\documentclass[12pt,oneside, a4paper]{article}

\usepackage{hyperref}	
\hypersetup{colorlinks,bookmarksopen,bookmarksnumbered,citecolor=blue,
linkcolor=black,pdfstartview=FitH,urlcolor=blue}

\usepackage{graphicx}
\usepackage{subfigure}
\usepackage{amssymb}
\usepackage{cite}
\usepackage{bm}
\usepackage{amsmath,amsthm}
\usepackage{cleveref}

\numberwithin{equation}{section}

\newcommand{\capdef}{}
\newcommand{\mycaption}[2][\capdef]{\renewcommand{\capdef}{#2}%
       \caption[#1]{{\footnotesize #2}}}

\begin{document}

\begin{titlepage}

\begin{center}

\vspace*{2cm}
        {\Large\bf Resolving the LMA-dark NSI degeneracy with coherent neutrino--nucleus scattering}
\vspace{1cm}

\renewcommand{\thefootnote}{\fnsymbol{footnote}}
{\bf Mariano Esteves Chaves}\footnote[1]{mchaves@ifi.unicamp.br}$^{1,2}$, 
{\bf Thomas Schwetz}\footnote[2]{schwetz@kit.edu}$^2$
\vspace{5mm}

{\it%
$^1${Instituto de F\'isica Gleb Wataghin, Universidade Estadual de Campinas (UNICAMP) \\13083-859, Campinas SP, Brazil}
}

{\it%
$^2$  {Institut f\"ur Astroteilchenphysik, Karlsruhe Institute of Technology (KIT),\\
    76021 Karlsruhe, Germany}
}

\vspace{8mm} 

\abstract{In the presence of non-standard neutrino interactions (NSI), a degeneracy exists in neutrino oscillation data, which involves the flipping of the octant of the mixing angle $\theta_{12}$ and the type of the neutrino mass ordering. In this article, we revisit the status of this degeneracy in the light of recent data on coherent elastic neutrino--nucleus scattering (CE$\nu$NS) from the COHERENT experiment. For general relative couplings to up and down quarks, the degeneracy is disfavoured at the $2\sigma$ level by the latest data but remains at a higher confidence level. We investigate the requirements of future CE$\nu$NS measurements to resolve the degeneracy with high significance. We find that a measurement involving both, electron and muon neutrino flavours and a target with a neutron-to-proton ratio close to 1 is required. For example, an experiment with a silicon target at the European Spallation Source can resolve the degeneracy at more than $4\sigma$ for arbitrary relative couplings to up and down quarks.}

\end{center}
\end{titlepage}

\renewcommand{\thefootnote}{\arabic{footnote}}
\setcounter{footnote}{0}

\setcounter{page}{2}
\tableofcontents

\section{Introduction}

Neutrino oscillation physics has entered the precision era, allowing
generic new physics to show up only as subleading corrections to the
well established standard three-flavour oscillation scenario
\cite{Esteban:2020cvm}. However, an exception to this statement is the
so-called LMA-dark degeneracy \cite{Miranda:2004nb} induced by
neutral-current (NC) type non-standard neutrino interactions
(NSI)~\cite{Wolfenstein:1977ue,Valle:1987gv,Guzzo:1991hi}.  The origin
of this degeneracy is a symmetry in the effective neutrino Hamiltonian
in matter \cite{GonzalezGarcia:2011my,Bakhti:2014pva}, which involves
flipping the octant of the mixing angle $\theta_{12}$ as well as a
sign change of the neutrino mass-squared difference $\Delta m^2_{31}$
\cite{Coloma:2016gei}. Therefore it is called also generalized
mass-ordering degeneracy. The degenerate solution is not a small
perturbation of the standard scenario and leads to a qualitatively
different mixing pattern. Since it is based on an exact symmetry of
the evolution equation it is fundamentally impossible to resolve the
degeneracy by any oscillation experiment or combination thereof, being
in vacuum or matter with an arbitrary profile.

The only way to resolve the generalized degeneracy is by
non-oscillation experiments
\cite{Escrihuela:2009up,Coloma:2016gei,Coloma:2017egw}, in particular,
neutral-current scattering experiments. A  promising candidate process
in this respect is coherent elastic neutrino--nucleus scattering
(CE$\nu$NS) \cite{Barranco:2005yy}, which has been recently observed for the first time by
the COHERENT experiment \cite{Akimov:2017ade}. Implications of this
first data for the LMA-dark degeneracy have been investigated by a
number of authors, e.g., \cite{Coloma:2017ncl,Liao:2017uzy,
  Denton:2018xmq, Giunti:2019xpr,Coloma:2019mbs,Miranda:2020tif,Dutta:2020che}. The authors of ref.~\cite{Coloma:2019mbs} performed a
combined analysis of global oscillation data with the COHERENT
CE$\nu$NS measurement on a CsI target reported in \cite{Akimov:2017ade}
in the framework of three-flavour oscillations
plus general NC NSI. The results of that work
show that the LMA-dark solution is ruled out at
more than $3\sigma$ if NSI occur exclusively with up or with down
quarks. However, allowing for general relative NSI strength to up and
down quarks, ref.~\cite{Coloma:2019mbs} reports a significant region
in parameter space where LMA-dark cannot be disfavoured with respect
to the standard mixing scenario. 


In this paper, we update the present status of LMA-dark by considering the recent measurement of CE$\nu$NS on argon \cite{Akimov:2020pdx} as well as a preliminary improved measurement with CsI by COHERENT~\cite{coherent2020_CsI_update}.
The LMA-dark solution becomes disfavoured with a $\Delta\chi^2 \approx 5$ relative to LMA-light, but the degeneracy persists at a higher confidence level. Then we investigate quantitatively under which conditions the LMA-dark degeneracy can be excluded (or eventually discovered) by future
CE$\nu$NS measurements, see \cite{raimund_strauss_2020_4124156} for a
review current and future projects. To be specific, our sensitivity studies will be motivated by possible CE$\nu$NS experiments at the European Spallation Source (ESS) \cite{Baxter:2019mcx} as well as by
two examples for projects at nuclear reactors, the ongoing CONNIE \cite{Aguilar-Arevalo:2019jlr} and CONUS
\cite{Bonet:2020awv} experiments.

Let us note that if NSI are induced by ``heavy'' mediator particles,
additional bounds apply, such as from high-energy scattering
experiments \cite{Dorenbosch:1986tb,Zeller:2001hh}, see e.g.,
\cite{Coloma:2017egw,Denton:2018xmq} for discussions, or (with some
model-dependence) also from
LHC~\cite{Friedland:2011za,Franzosi:2015wha}. Here we assume that the
contact interaction approximation is valid only up to the momentum
transfers as relevant for coherent scattering experiments at stopped
pion sources, i.e., of order $50$~MeV, and we do not take
into account information from data which would require contact
interactions to be valid also at higher energy scales.

The outline of the paper is as follows. In section~\ref{sec:nsi} we introduce the NSI formalism and fix our notation. Then we review the LMA-dark degeneracy as well as the parametric dependencies of CE$\nu$NS measurements. In section~\ref{sec:present} we discuss the present status of the LMA-degeneracy from a combination of data from neutrino oscillations with the COHERENT experiment, including the recent measurement of CE$\nu$NS on argon and CsI updates. In section~\ref{sec:future} we discuss the sensitivity of future measurements based on neutrinos from a stopped pion source (sec.~\ref{sec:ESS}) and from nuclear reactors (sec.~\ref{sec:reactors}). We summarize our results in section~\ref{sec:conclusion}.

\section{NSI formalism and the LMA-dark degeneracy}
\label{sec:nsi}

In this section, we introduce the formalism for the NSI considered in
this paper. They are described by an effective dim-6 interaction
Lagrangian, in analogy to the 4-Fermi interaction.  We follow closely
the notation of Refs.~\cite{Esteban:2018ppq,Coloma:2019mbs}, where
also the latest bounds from a global analysis are presented.  Recent
reviews on NSI can be found in
Refs.~\cite{Miranda:2015dra,Farzan:2017xzy,Dev:2019anc}.

We consider NSI of the NC type with up- and down-quarks in the
background medium, which are described by the effective 
Lagrangian
\begin{equation}\label{eq:L}
  \mathcal{L}_{\rm NSI} = -2\sqrt{2} G_F
  \sum_{\alpha, \beta} \sum_{f=u,d}
  \epsilon_{\alpha\beta}^f
  (\overline\nu_{\alpha L} \gamma^\mu \nu_{\beta L}) (\overline f \gamma_\mu f) \,,
\end{equation}
where, $\alpha, \beta = e,\mu,\tau$. The dimensionless parameters
$\epsilon_{\alpha\beta}^f$ parametrize the strength of the new
interaction with respect to the Fermi constant $G_F$. Hermiticity
requires that $\epsilon_{\alpha\beta}^f =
(\epsilon_{\beta\alpha}^f)^*$. Note that we consider only vector
interactions since we are interested in the contribution to the
effective matter potential. The embedding of the low-energy effective
interaction of eq.~\eqref{eq:L} into a consistent effective field
theory in the framework of the Standard Model has been discussed in
\cite{Altmannshofer:2018xyo,Falkowski:2019kfn,Bischer:2019ttk}.

Following Ref.~\cite{Esteban:2018ppq} we allow for arbitrary relative
couplings to up- and down-quarks parametrized by a parameter $\eta$,
but we assume that the relative up- and down-quark coupling
is independent of the neutrino flavour:
\begin{equation}\label{eq:eps_eta}
  \epsilon^u_{\alpha\beta} = \frac{\sqrt{5}}{3}(2\cos\eta -\sin\eta)\epsilon^\eta_{\alpha\beta}
  \,,\qquad
  \epsilon^d_{\alpha\beta} = \frac{\sqrt{5}}{3}(2\sin\eta -\cos\eta)\epsilon^\eta_{\alpha\beta}
  \,.
\end{equation}
The analysis is performed in terms of the coefficients
$\epsilon^\eta_{\alpha\beta}$ and the angle $\eta$.
The normalization is chosen such that
$\epsilon^\eta_{\alpha\beta} = \epsilon^{u}_{\alpha\beta}$, 
$\epsilon^{d}_{\alpha\beta} = 0$ 
for $\eta = \arctan(1/2) \approx 26.6^\circ$
and
$\epsilon^\eta_{\alpha\beta} = \epsilon^{d}_{\alpha\beta}$, 
$\epsilon^{u}_{\alpha\beta} = 0$ 
for $\eta = \arctan(2) \approx 63.4^\circ$.
The effective couplings to protons and neutrons are obtained as
\begin{equation}
  \epsilon^p_{\alpha\beta} =
  2\epsilon^{u}_{\alpha\beta}  +\epsilon^d_{\alpha\beta} =
  \sqrt{5} \epsilon^\eta_{\alpha\beta} \cos\eta
  \,,\qquad 
  \epsilon^n_{\alpha\beta} =
  2\epsilon^d_{\alpha\beta}  +\epsilon^u_{\alpha\beta} =
  \sqrt{5} \epsilon^\eta_{\alpha\beta}\sin\eta
  \,.
\end{equation}

These NSI will contribute to the effective matter
potential~\cite{Wolfenstein:1977ue} in the Hamiltonian relevant for
neutrino propagation. Since the flavour evolution is only sensitive to
phase differences, oscillations are sensitive only to two differences of
flavour diagonal NSI, for instance $\epsilon^{\eta}_{ee} -
\epsilon^{\eta}_{\mu\mu}$ and $\epsilon^{\eta}_{\mu\mu} -
\epsilon^{\eta}_{\tau\tau}$, as well as the three (complex)
off-diagonal coefficients $\epsilon^{\eta}_{\alpha\beta}
\,(\alpha\neq\beta)$.  Neutrino scattering experiments are sensitive also to
the individual diagonal NSI, $\epsilon^{\eta}_{\alpha\alpha}$.

Let us consider now a NC-type interaction with a medium consisting of nuclei with $Z$ protons and $N$ neutrons. We define an effective NSI parameter depending on the neutron-to-proton ratio $Y=N/Z$ by
\begin{equation}\label{eq:epsY}
  \epsilon^{Y,\eta}_{\alpha\beta} =  \epsilon^p_{\alpha\beta} + Y \epsilon^n_{\alpha\beta}
  = \sqrt{5} \epsilon^\eta_{\alpha\beta}(\cos\eta + Y \sin\eta)\,.
\end{equation}
Note that in general $\epsilon^{Y,\eta}_{\alpha\beta}$ may change along a
given neutrino trajectory, if the neutron-to-proton ratio changes
along the path. Only if $\epsilon^n_{\alpha\beta} = 0$, i.e., for
$\eta=0$, $\epsilon^{Y,\eta}_{\alpha\beta}$ becomes position independent and
is equal to the NSI with protons.

\subsection{The LMA-dark degeneracy}

Due to CPT invariance, neutrino evolution is unchanged when the
effective Hamiltonian is transformed as $H \to -H^*$. For the Hamiltonian relevant for neutrino propagation in vacuum, this transformation can be implemented by changing the parameters as
\begin{align}\label{eq:trf-vac}
  \Delta m^2_{31} \to -\Delta m^2_{32} \,,\qquad
  \theta_{12} \to \pi/2 - \theta_{12} \,,\qquad
  \delta_{\rm CP} \to \pi - \delta_{\rm CP} \,,
\end{align}
see \cite{Coloma:2016gei} for the parameterisation
conventions.  The standard matter effect breaks this degeneracy, which
allows one to fix the sign of $\Delta m^2_{31}$ (i.e., the mass
ordering) as well as the octant of $\theta_{12}$ by observing the
standard matter effect. However, if we allow for NSI, the symmetry can
be restored by performing simultaneously to eq.~\eqref{eq:trf-vac} the
transformation
\cite{GonzalezGarcia:2011my,Gonzalez-Garcia:2013usa,Bakhti:2014pva,Coloma:2016gei}
\begin{align}
  (\epsilon^{Y,\eta}_{ee} - \epsilon^{Y,\eta}_{\mu\mu}) &\to 
  - (\epsilon^{Y,\eta}_{ee} - \epsilon^{Y,\eta}_{\mu\mu}) - 2 \,, \nonumber\\ 
  (\epsilon^{Y,\eta}_{\tau\tau} - \epsilon^{Y,\eta}_{\mu\mu}) &\to 
  - (\epsilon^{Y,\eta}_{\tau\tau} - \epsilon^{Y,\eta}_{\mu\mu}) \,, \label{eq:trf-nsi}\\ 
  \epsilon^{Y,\eta}_{\alpha\beta} &\to - (\epsilon^{Y,\eta}_{\alpha\beta})^* \qquad (\alpha\neq\beta) \,. \nonumber
\end{align}
Hence, if NSI parameters which can accommodate this transformation are
allowed by non-oscillation data, neither the octant of the mixing angle $\theta_{12}$ \cite{Miranda:2004nb} nor the neutrino mass ordering \cite{Coloma:2016gei} can be determined because of the degeneracy.  As mentioned above, for $\eta=0$ (NSI with protons only),
the parameters $\epsilon^{Y,\eta}_{\alpha\beta}$ are independent of the
chemical composition of the background medium, and therefore the
degeneracy is exact and holds for any combination of oscillation experiments including arbitrary matter density profiles. For other values of $\eta$, the degeneracy is
still present approximately for the actually available data.

A detailed investigation of the status of the LMA-dark degeneracy has
been presented recently in Ref.~\cite{Coloma:2019mbs} by performing a
global analysis of neutrino oscillation data combined with data from
the COHERENT CE$\nu$NS measurement. While the SM fit with no NSI
provides a good fit to the data, the authors find that in the region
$-40^\circ \lesssim \eta \lesssim 0^\circ$ the LMA-dark degeneracy
cannot be excluded relative to the SM fit (the precise range depends on details of the COHERENT data analysis).  The main purpose of this paper is to study the requirements which are needed to
resolve the degeneracy in this range of $\eta$. The size of NSI parameters is roughly $1 \lesssim (\epsilon^\eta_{\mu\mu}-\epsilon^\eta_{ee}) 
\lesssim 2$, $\epsilon^\eta_{\tau\tau} \approx
\epsilon^\eta_{\mu\mu}$, and the off-diagonal coefficients consistent
with zero. Hence, as suggested by the first line in
eq.~\eqref{eq:trf-nsi}, the LMA-dark solution implies NSI coefficients
of order one.

\subsection{CE$\nu$NS}

The differential cross section for coherent scattering of a neutrino
with energy $E_\nu$ on a 
nucleus with $Z$ protons, $N$ neutrons, and mass $M$
reads \cite{Freedman:1973yd}:
\begin{align}\label{eq:sigma-coh}
  \frac{d\sigma}{dT} = \frac{G_F^2}{2\pi} Q^2 F^2(q^2)M\left(2-\frac{MT}{E_\nu^2}\right) \,.
\end{align}
Here, $T$ is the recoil energy of the nucleus, $F(q^2)$ is the
nuclear form factor depending on the momentum transfer squared, $q^2 =
2 MT$ and $Q^2$ is the weak charge of the nucleus. In the Standard Model (SM) its value is
\begin{align}\label{eq:QSM}
  Q_{\rm SM}^2 = (Zg_p^V + N g_n^V)^2 \,,
\end{align}
with the tree-level relations $g_p^V = 1/2 - 2 \sin^2\theta_W$ and $g_n^V=-1/2$. We have checked that loop-corrections to these values (see e.g., \cite{Cadeddu:2020lky} for a discussion) have a negligible impact on our numerical results; therefore, for the sake of simplicity, we stick to the tree-level values for $Q_{\rm SM}$. For the weak mixing angle $\theta_W$, we adopt the low-energy value derived in \cite{Erler:2004in}. 

The effect of NC vector NSI as considered in this work can be taken into account by replacing $Q_{\rm SM}^2$ by an effective weak charge, which now becomes dependent on the flavour $\alpha$ of the incoming neutrino \cite{Barranco:2005yy}:
\begin{align}
  Q_{\alpha}^2 &= \left[Z(g_p^V+\epsilon^p_{\alpha\alpha}) + N (g_n^V+\epsilon^n_{\alpha\alpha})\right]^2
  + \sum_{\beta\neq\alpha}
  \left[Z \epsilon^p_{\alpha\beta} + N \epsilon^n_{\alpha\beta}\right]^2
  \nonumber\\
  &= \left( Q_{\rm SM} + Z\epsilon^{Y,\eta}_{\alpha\alpha} \right)^2 + 
  Z^2 \sum_{\beta\neq\alpha} \left(\epsilon^{Y,\eta}_{\alpha\beta}\right)^2 \,,
\label{eq:QNSI}
\end{align}
with $\epsilon^{Y,\eta}_{\alpha\beta}$ defined in eq.~\eqref{eq:epsY} and for
the sake of simplicity, here and in the following we assume that
off-diagonal NSI coefficients are real. The first term in
eq.~\eqref{eq:QNSI} corresponds to the flavour diagonal process
$\nu_\alpha + A \to \nu_\alpha + A$, where the NSI induced amplitude
can interfere with the SM term, whereas the second term corresponds to
flavour changing scattering, $\nu_\alpha + A \to \nu_\beta + A$.

It is clear from eqs.~\eqref{eq:QNSI} and \eqref{eq:epsY} that an experiment with a given target nucleus will not be sensitive to NSI if $\epsilon_{\alpha\beta}^{Y,\eta}=0$ because of the negative interference of the interactions with protons and neutrons, which happens for 
\begin{align}\label{eq:eta-blind}
  \eta_{\rm blind} = - \arctan\left(\frac{1}{Y}\right) \,.
\end{align}
In Tab.~\ref{tab:elements} we list several of possible detector targets showing their respective values for $Z$, $Y$, and $\eta_{\rm blind}$.
LMA-dark is allowed by present data for values of $\eta$ somewhat larger---but close to---the blind spot for CsI, $\eta_{\rm blind}^{\rm CsI} \approx -35.4^\circ$. In order to resolve the degeneracy, data from a target with an 
$\eta_{\rm blind}$ sufficiently smaller than this value
will be needed.

\begin{table}[]
\centering
\begin{tabular}{c|cccccc}
Target & $Z$ & $Y$ & $\eta_{\rm blind}$ & $-Q_{\rm SM}$ &  $\sigma/Q^2_{\rm SM}$ & $\sigma_{\mu}/\sigma$
\\ \hline
C$_3$F$_8$ & 8.2& 1.081 & $-42.8^\circ$ &  4.27 & $13.3\%$ & $\infty$\\
Si         & 14 & 1.006 & $-44.8^\circ$ &  6.72 & $17.6\%$ & $\infty$\\
Ar         & 18 & 1.235 & $-39.0^\circ$ & 10.71 & $12.0\%$ & $\infty$\\
Ge         & 32 & 1.270 & $-38.2^\circ$ & 19.6 & $14.2\%$ & 4.20\\
CsI        & 54 & 1.405 & $-35.4^\circ$ & 36.7 & $12.5\%$ & 3.37\\
Xe         & 54 & 1.431 & $-35.0^\circ$ & 37.4 & $12.0\%$ & 4.01
\end{tabular}
\mycaption{Number of protons $Z$, the neutron-to-proton ratio $Y=N/Z$, the corresponding blind spot $\eta_{\rm blind}$, eq.~\eqref{eq:eta-blind}, and the value of the SM weak charge, $Q_{\rm SM}$, for different target materials. We use the average $N$ corresponding to the natural isotope abundances, and for the molecules C$_3$F$_8$ and CsI we take the average $Z$ and $N$ values. The last two columns show our assumptions about the measurement uncertainties obtainable at ESS, see eq.~\eqref{eq:chi-ess}.}
\label{tab:elements} 
\end{table}

\section{Present status of the LMA-dark degeneracy}
\label{sec:present}

\subsection{Analysis details}

{\bf Statistical method.} In order to test the LMA-dark solution we define the following $\Delta\chi^2$ function:
\begin{align}\label{eq:Dchisq-osc}
    \Delta\chi^2(\epsilon^\eta,\eta) = 
    \chi^2_D(\epsilon^\eta,\eta) -
    \chi^2_{L,{\rm min}} \,,\qquad
    \chi^2_{L,{\rm min}} =
    \min_{\epsilon^\eta,\eta}
    \chi^2_L(\epsilon^\eta,\eta) \,.     
\end{align}
Here, $\epsilon^\eta = (\epsilon^\eta_{\alpha\beta})$ is a short-hand for all NSI coefficients and $\chi^2_{D,L}(\epsilon^\eta,\eta)$ are the $\chi^2$ functions restricted to the LMA-dark ($\theta_{12} > 45^\circ$) or LMA-light ($\theta_{12} < 45^\circ$) sides of the parameter space, respectively. 
The $\Delta\chi^2$ in eq.~\eqref{eq:Dchisq-osc} quantifies the degree to which the LMA-dark solution is disfavoured with respect to LMA-light. In particular, 
\begin{align}\label{eq:Dchisq-min} 
    \Delta\chi^2_{DL} \equiv
    \min_{\epsilon^\eta,\eta}\left[
    \Delta\chi^2(\epsilon^\eta,\eta)\right] = 
    \chi^2_{D,{\rm min}} -
    \chi^2_{L,{\rm min}}  
\end{align}
corresponds to the log-likelihood ratio of the two hypotheses LMA-dark versus LMA-light. In the following, we will evaluate $\Delta\chi^2_{DL}$ from eq.~\eqref{eq:Dchisq-min} for 1~dof to quantify the exclusion of the LMA-dark degeneracy. In order to perform the minimisation over the NSI parameters, 
we use a Monte Carlo minimisation based on the differential evolution method \cite{differential_evolution}. 
The numerical calculations are performed with the SciPy library \cite{2020SciPy-NMeth} in python that already has this algorithm implemented.

\bigskip
{\bf Oscillation experiments.} The information from oscillation experiments is included by using the results of the global analysis from ref.~\cite{Esteban:2018ppq}. We re-construct approximate functions $\chi^2_{L,{\rm osc}}(\epsilon^\eta, \eta)$ and $\chi^2_{D,{\rm osc}}(\epsilon^\eta, \eta)$ from figs.~7 and 10 of ref.~\cite{Esteban:2018ppq} (2020 updated version).\footnote{We are grateful to the authors of ref.~\cite{Esteban:2018ppq} for providing us a $\chi^2$-table corresponding to an updated version of their fig.~7. Let us note that the figure shows marginalized regions for each $\epsilon^\eta_{\alpha\beta}$ as a function of $\eta$. Therefore, we neglect correlations between the different $\epsilon^\eta_{\alpha\beta}$. In general, our exclusions will be conservative; if parameter correlations can be included exclusions of the LMA-dark degeneracy would be somewhat stronger.} We refer the reader to that reference for further details about the statistical analysis and used data. 

\begin{figure}
    \centering
    \includegraphics{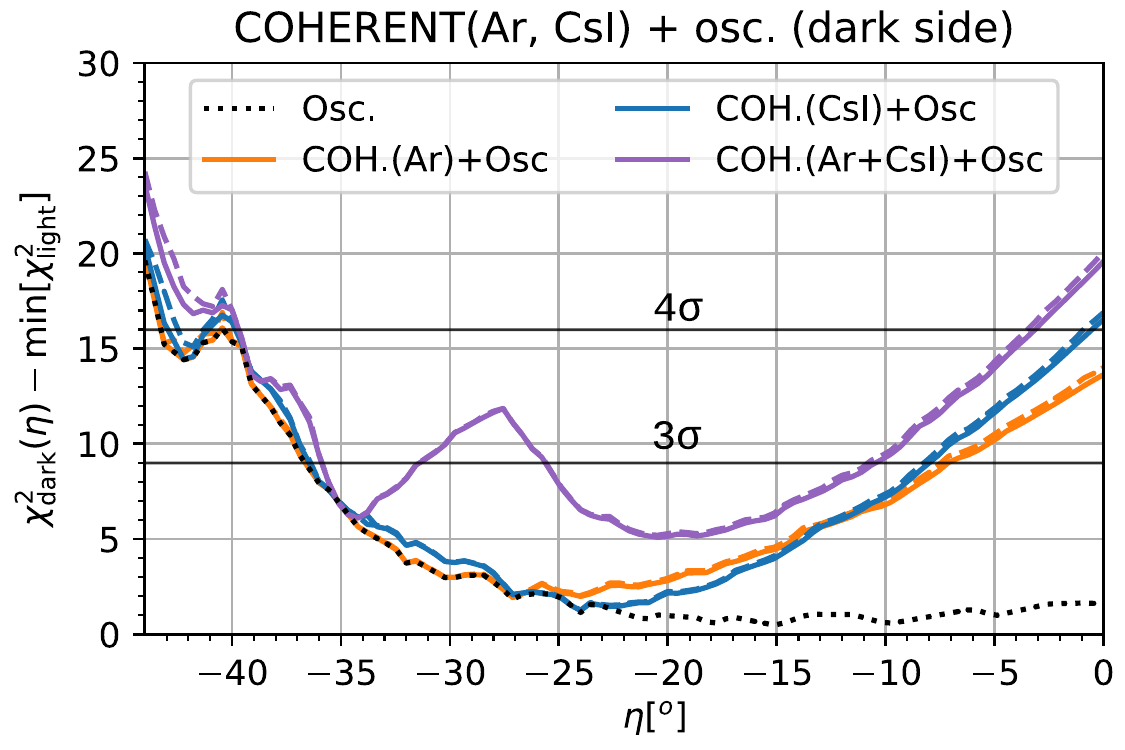}
    \mycaption{$\Delta\chi^2$ of the LMA-dark solution minimized with respect to $\epsilon_{\alpha\beta}^\eta$ as a function of $\eta$ for oscillation data (black), oscillations+Coherent(CsI) (blue), oscillations+Coherent(Ar) (orange), and all three data sets combined (purple). For dashed curves the off-diagonal 
    $\epsilon_{\alpha\beta}^\eta$ are fixed at zero, for solid curves we minimize with respect to them.  }
    \label{fig:eta-coh-osc}
\end{figure}

The function $\Delta\chi^2_{\rm osc}$ minimized with respect to all $\epsilon^\eta$ is shown as black-dotted curve in fig.~\ref{fig:eta-coh-osc}. We restrict to the range $-50^\circ \lesssim \eta \lesssim 0^\circ$, since outside this region LMA-dark is strongly disfavoured~\cite{Coloma:2019mbs}. 
We see that oscillation data by themselves exclude LMA-dark for values of $\eta \lesssim -37^\circ$ at more than $3\sigma$,
while for $-25^\circ < \eta < 0^\circ$ LMA-dark provides a comparable fit as LMA-light with $\Delta\chi^2_{DL} \lesssim 2$.

\bigskip

{\bf COHERENT.} Let us now discuss our implementation of CE$\nu$NS data from the COHERENT experiment~\cite{Akimov:2017ade}. They are using neutrinos from a stopped pion source, consisting of $\nu_\mu$ and $\nu_e$ flavours in the ratio 2:1. Therefore, this experiment is sensitive to both, $Q_\mu$ and $Q_e$ weak charges, see eq.~\eqref{eq:QNSI}. Via a combined fit of the event energy spectrum and time distribution, it is possible to separate the contribution of the $\mu$ and $e$ flavours statistically to some extent. 

The original data \cite{Akimov:2017ade}
is based on a CsI target, and their implications for NSI have been widely studied, see e.g., \cite{Coloma:2017ncl,Liao:2017uzy,Denton:2018xmq,Giunti:2019xpr,Miranda:2020tif,Dutta:2020che,Coloma:2019mbs}. Here we update the CsI analysis by using the preliminary results presented in ref.~\cite{coherent2020_CsI_update}. Compared to the original 2017 data \cite{Akimov:2017ade}, the statistics is increased, and---most importantly---new data on the quenching factor have become available, leading to an over-all improvement in precision from 33\% of the 2017 analysis to about 16\%. 

Ref.~\cite{coherent2020_CsI_update} reports results in terms of the correlated determination of averaged cross sections $\langle\sigma_\mu\rangle$ and
$\langle\sigma_e\rangle$, corresponding to the $\nu_\mu$ and $\nu_e$ flux averaged contributions to the observed CE$\nu$NS cross section. Using that $\langle\sigma_\alpha\rangle \propto Q_\alpha^2$ ($\alpha=e,\mu$) we can reconstruct a $\chi^2$ function by
\begin{equation}
    \chi_{\text{Coh(CsI)}}^2(Q_e^2,Q_\mu^2) = 
    \left(\Delta Q_e^2 , \, \Delta Q_\mu^2 \right)
    \begin{pmatrix}\sigma_e^2 & \rho \sigma_e\sigma_\mu\\ \rho \sigma_e\sigma_\mu& \sigma_\mu^2\end{pmatrix}^{-1}\begin{matrix}\begin{pmatrix}\Delta Q_e^2 \\ \Delta Q_\mu^2 \end{pmatrix}&\mbox{}\end{matrix},
    \label{eq:chi-coh-CsI}
\end{equation}
where $\Delta Q_\alpha^2 = Q_\alpha^2-(Q_{\alpha}^\text{bfp})^2$. 
With the values $\rho = -0.790$, $\sigma_e=1204.7$, $\sigma_\mu=404.6$, $(Q_e^\text{bfp})^2 = 1200.0$ and $(Q_\mu^\text{bfp})^2 = 1245.1$ we can reproduce the results shown in  \cite{coherent2020_CsI_update} rather accurately. Using eq.~\eqref{eq:QNSI} it is straight forward to transform $\chi_{\text{Coh(CsI)}}^2(Q_e^2,Q_\mu^2)$
into $\chi_{\text{Coh(CsI)}}^2(\epsilon^\eta, \eta)$.

In addition to CsI data, COHERENT has also published a first measurements of CE$\nu$NS on an argon target~\cite{Akimov:2020pdx}. Implications for various new-physics searches of the Ar data have been investigated e.g., in~\cite{Cadeddu:2020lky,Miranda:2020tif}.
We estimate the weak charge determination from these data by assuming that the measurement is dominated by the total rate. Following ref.~\cite{Coloma:2017ncl}, we adopt the following $\chi^2$ definition:
\begin{equation}
    \chi^2_{\text{Coh(Ar)}}(Q_e^2,Q_\mu^2) = \frac{\left[f_eQ_e^2+f_\mu Q_\mu^2-(Q^{\text{bfp}})^2\right]^2}{\sigma^2} \,.
    \label{eq:chi-coh-Ar}
\end{equation}
For the relative contributions of $\nu_e$ and $\nu_\mu$ flavours we adopt values similar to the ones for CsI data, $f_e\approx0.3$ and $f_\mu\approx0.7$~\cite{Akimov:2017ade,Coloma:2017ncl}.
The $\sigma=25.0$ is the total rate error and $Q^{\text{bfp}}=-12.2$ is the best-fit point measured. These values are estimated by reproducing fig.~6 of ref.~\cite{Akimov:2020pdx}. We combine Ar and CsI measurements from COHERENT assuming no correlation between the measurements. 

\begin{figure}[t]
    \centering
    \includegraphics[scale=0.9]{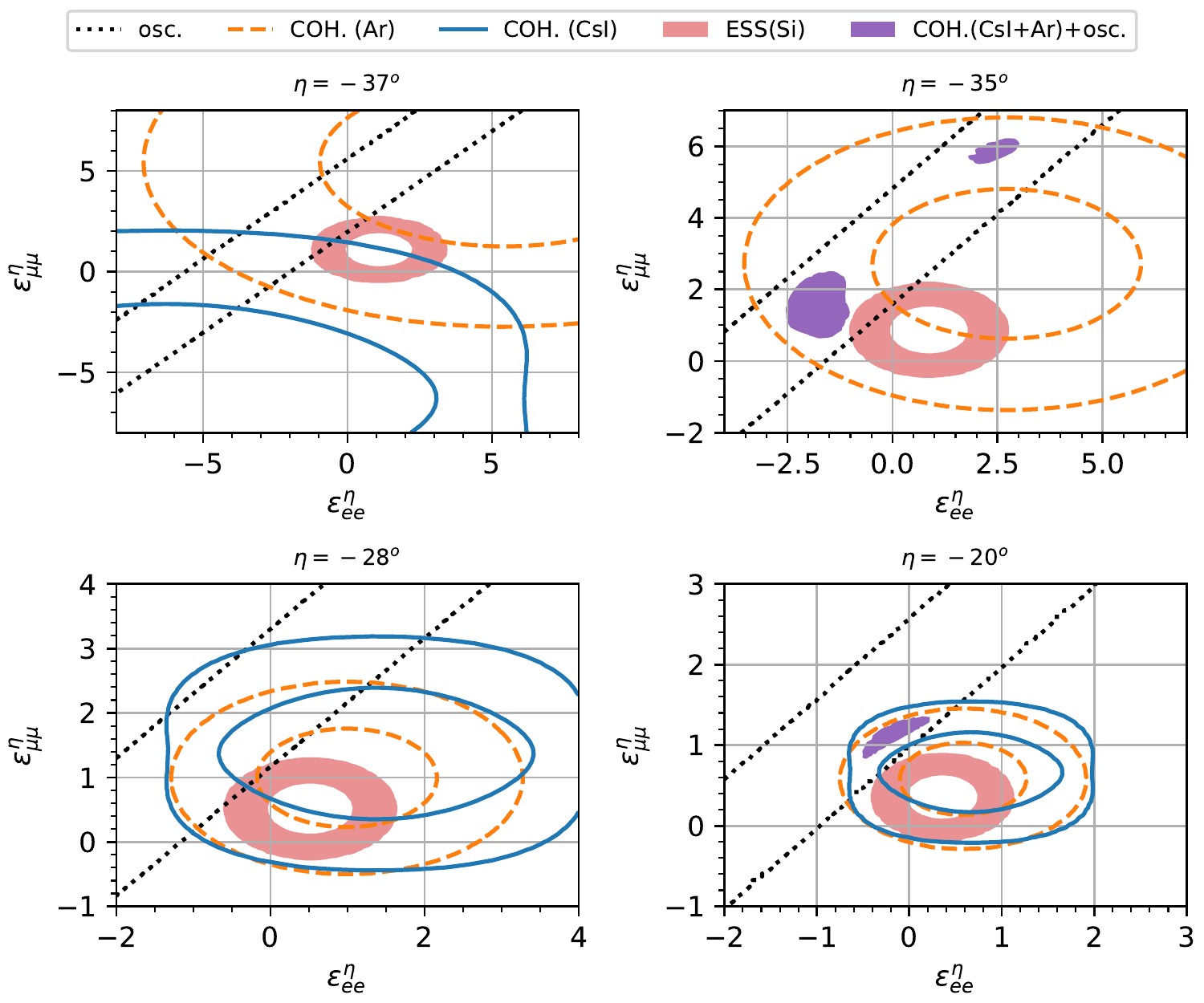}
    \mycaption{Allowed regions in the plane of $\epsilon_{ee}^\eta$ and $\epsilon_{\mu\mu}^\eta$ at $\Delta\chi^2=11.83$ corresponding to $3\sigma$ for 2~dof, for different fixed values of $\eta$. The contour lines correspond to the regions from oscillation data, Coh(CsI), and Coh(Ar) separately. The purple filled region is obtained by combining all three data sets. The light-red filled region shows the sensitivity of a future measurement at ESS using a Si detector.
    Off-diagonal $\epsilon^\eta_{\alpha\beta}$ are fixed at zero.}
    \label{fig:contours}
\end{figure}

\subsection{Results from current data}

The results from combining oscillation data with COHERENT CsI and/or Ar are summarized in Figs.~\ref{fig:eta-coh-osc} and \ref{fig:contours}. 
For each data combination, we construct $\chi^2_{L,D}$ and build the $\Delta\chi^2$ between dark and light sides according to eq.~\eqref{eq:Dchisq-osc}, on which the figures are based.
When combining all data, LMA-dark becomes disfavoured for all possible values of $\eta$ with $\Delta\chi^2 > 5$, i.e., at the $2.2\sigma$ level, see purple curve in fig.~\ref{fig:eta-coh-osc}. However, the degeneracy remains below $3\sigma$ for $\eta$ in the intervals $[-35.9^\circ,-31.3^\circ]$ and $[-25.6^\circ,-10.5^\circ]$

We note that the interplay of the three data sets, oscillations, Coh(CsI), Coh(Ar) is essential for this result. By using only Coh(CsI) or Coh(Ar) data, LMA-dark remains allowed below $2\sigma$ in the region $-31^\circ \lesssim \eta \lesssim -14^\circ$. 
In fig.~\ref{fig:contours} we show the individual constraints in the plane of $\epsilon_{ee}^\eta$ and $\epsilon_{\mu\mu}^\eta$ for different fixed values of $\eta$. In the left panels corresponding to $\eta=-37^\circ$ and $-28^\circ$ the complementarity of CsI and Ar data pushes the $\Delta\chi^2$ above the $3\sigma$ level and therefore no allowed regions appear for the combined data. The right panels correspond to the minima of $\Delta\chi^2$ around $\eta=-35^\circ$ and $-20^\circ$, illustrating why present data cannot disfavour these regions. The upper right panel corresponds to the blind spot for CsI and therefore no limits are visible for Coh(CsI).

\begin{figure}[t]
    \centering
    \includegraphics[scale=0.9]{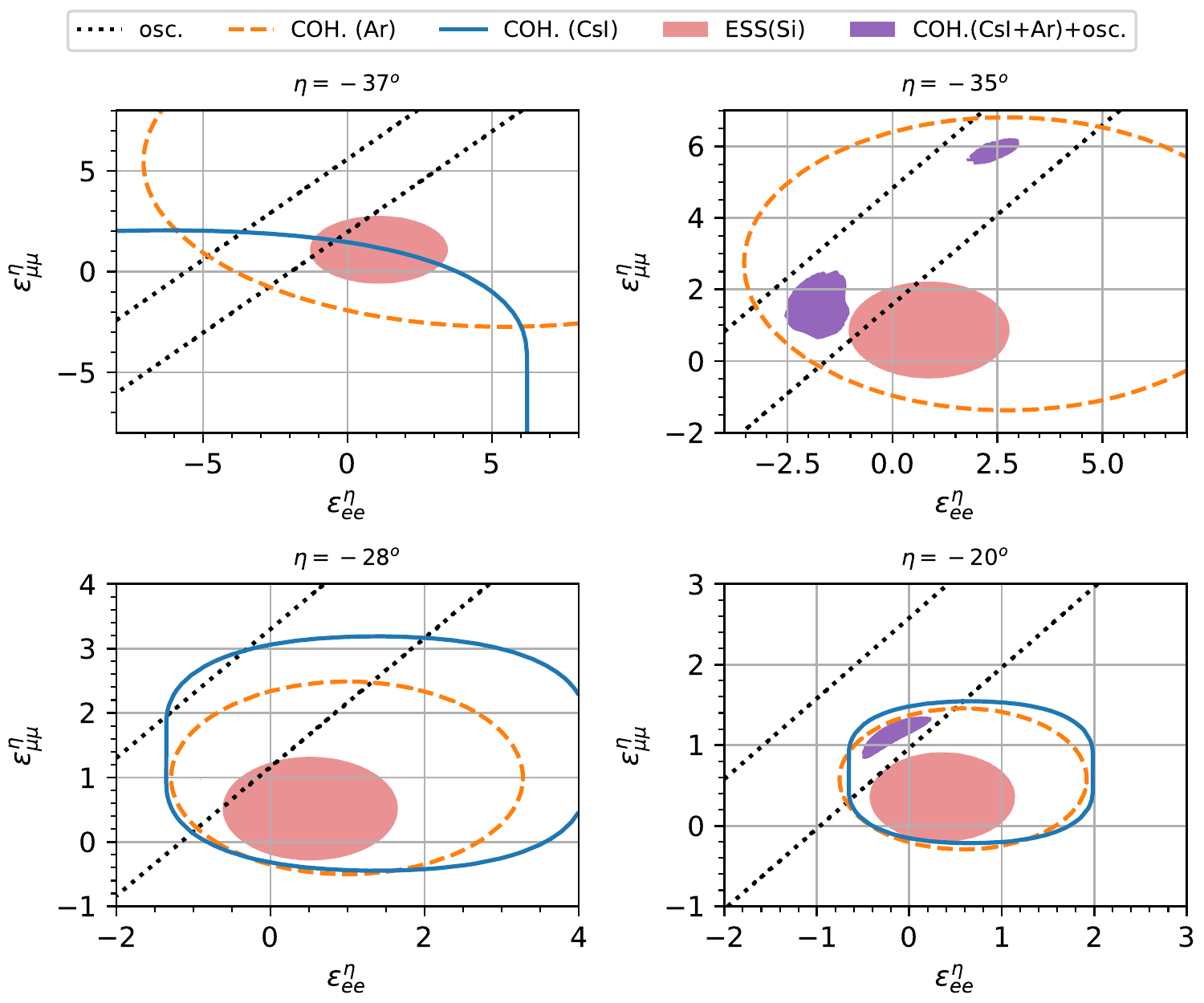}
    \mycaption{Same as fig.~\ref{fig:contours} but minimizing with respect to the off-diagonal $\epsilon^\eta_{\alpha\beta}$.}
    \label{fig:contours-off}
\end{figure}

In fig.~\ref{fig:contours} we set all off-diagonal NSI parameters $\epsilon^\eta_{\alpha\beta}$ with $\alpha\neq\beta$ to zero. Fig.~\ref{fig:contours-off} shows the effect when allowing them to vary freely and minimizing the $\chi^2$ with respect to them. We see that this leads to the disappearance of the ``hole'' in the allowed regions from COHERENT. This can be easily understood from considering eq.~\eqref{eq:QNSI}. The hole appears due to a negative interference of the diagonal NSI with $Q_{\rm SM}$, such that $(Q_{\rm SM} + Z\epsilon^{Y,\eta}_{\alpha\alpha})^2 < Q_{\rm SM}^2$. This can be compensated by the off-diagonal NSIs in the second term in eq.~\eqref{eq:QNSI}. 

However, in order to fill the holes in the COHERENT regions, relatively large values of off-diagonal $\epsilon^\eta_{\alpha\beta}$ are required, whereas oscillation data puts rather strong bounds on them. As visible from 
fig.~7 of ref.~\cite{Esteban:2018ppq}, the weakest bound is 
$\epsilon^\eta_{e\tau} \lesssim 0.2$ at $3\sigma$ and significantly tighter bounds on $\epsilon^\eta_{e\mu}$ and $\epsilon^\eta_{\mu\tau}$.
Once we restrict the off-diagonal terms to their allowed range by including the $\chi^2$ corresponding to the respective panels in fig.~7 of ref.~\cite{Esteban:2018ppq}, their effect is negligible and we recover the situation shown in fig.~\ref{fig:contours}. This can also be seen in fig.~\ref{fig:eta-coh-osc} by comparing dashed curves (off-diagonal NSI fixed to zero) and solid curves (minimized with respect to off-diagonal NSI).

From the purple regions in fig.~\ref{fig:contours-off} we read off the NSI parameters, for which the LMA-dark degeneracy is allowed below $3\sigma$ from current data. The three islands are located roughly around
\begin{align}\label{eq:eps-LMA-D}
\begin{array}{lll}
    \eta \approx -20^\circ \,,\quad 
    &\epsilon^\eta_{ee} \approx -0.2 \,,\quad 
    &\epsilon^\eta_{\mu\mu} \approx 1.2 \,, 
    \\
    \eta \approx -35^\circ \,,\quad 
    &\epsilon^\eta_{ee} \approx -1.8 \,,\quad 
    &\epsilon^\eta_{\mu\mu} \approx 1.5 \,,
    \\
    \eta \approx -35^\circ \,,\quad 
    &\epsilon^\eta_{ee} \approx 2.5 \,,\quad 
    &\epsilon^\eta_{\mu\mu} \approx 6.0 \,.
\end{array}
\end{align}

\section{Resolving the degeneracy with future CE$\nu$NS data}
\label{sec:future}

Let us now investigate which future CE$\nu$NS measurements have the potential to exclude LMA-dark with high significance. We build on and extend the results from \cite{Coloma:2017egw}, where related discussions can be found.
We consider two examples, namely CE$\nu$NS measurements using either stopped pions or a nuclear reactor as neutrino sources. Apart from the different neutrino energies, the main difference for our purposes is that stopped pion sources produce a mixture containing electron and muon (anti-)neutrino flavours, whereas nuclear reactors are a source of pure electron anti-neutrinos. Below we will always assume that the best-fit point for a hypothetical future experiment is at $Q_{\rm SM}^2$, i.e., no NSI. Then we calculate the sensitivity to constrain $Q_e^2$ and $Q_\mu^2$ under some assumptions about the measurement uncertainty and add the resulting $\chi^2$ to the one from present data as discussed in the previous section.

\subsection{CE$\nu$NS from a stopped pion neutrino source}
\label{sec:ESS}

To be specific, in this section we consider as an example for a stopped pion source the sensitivity of a possible  CE$\nu$NS measurement at the European Spallation Source (ESS)~\cite{Baxter:2019mcx}. The ESS can provide an increase in neutron luminosity by a factor 30--100 with respect to previous spallation sources, and an order of magnitude larger neutrino fluxes than the SNS where the COHERENT experiment is located. The sensitivity of CE$\nu$NS measurements using different detector technologies based on various target materials has been investigated in ref.~\cite{Baxter:2019mcx}, where details about the assumed experimental configurations can be found. See also \cite{Miranda:2020syh} for some physics applications. 

The CE$\nu$NS measurement at ESS will be dominated by the total rate of the signal. We adopt the neutrino flavour contribution to the event rate of $\nu_e:\nu_\mu:\overline{\nu}_\mu = (1:1:1)$ and therefore the measurement corresponds to the determination of the effective weak charge combination $Q_e^2/3+2Q_\mu^2/3$. In some cases the detector energy resolution allows to partially distinguish between electron neutrinos and muon neutrinos due to the different spectral shape of their respective fluxes~\cite{Baxter:2019mcx}. To implement this effect we suppose that an additional independent constraint on $Q_\mu$ can be obtained. Hence, we use the following $\chi^2$ definition:
\begin{equation}
    \chi^2_{\rm ESS}=\frac{(Q_{\rm SM}^2-Q_e^2/3-2Q_\mu^2/3)^2}{\sigma^2} +
    \frac{(Q^2_{\rm SM}-Q_\mu^2)^2}{\sigma_\mu^2} \,.
    \label{eq:chi-ess}
\end{equation}
Here, $\sigma$ ($\sigma_\mu$) is the assumed uncertainty on the total rate (on $Q^2_{\mu}$). The values of the uncertainties have been chosen in order to match fig.~12 of \cite{Baxter:2019mcx} and the numbers are listed for the various target materials in tab.~\ref{tab:elements}. In good agreement with the assumptions from \cite{Baxter:2019mcx}, we find rate measurement uncertainties in the range from 12\% to 18\%. For Ge, CsI, and Xe targets a constraint on $Q^2_\mu$ with $\sigma_\mu/\sigma \simeq 4$ can be obtained, whereas for the lighter targets C$_3$F$_8$, Si, Ar the measurement is dominated by the total rate alone. With these assumptions we can reproduce fig.~12 of~\cite{Baxter:2019mcx} with excellent accuracy.

\begin{figure}[t]
    \centering
    \includegraphics[scale=1]{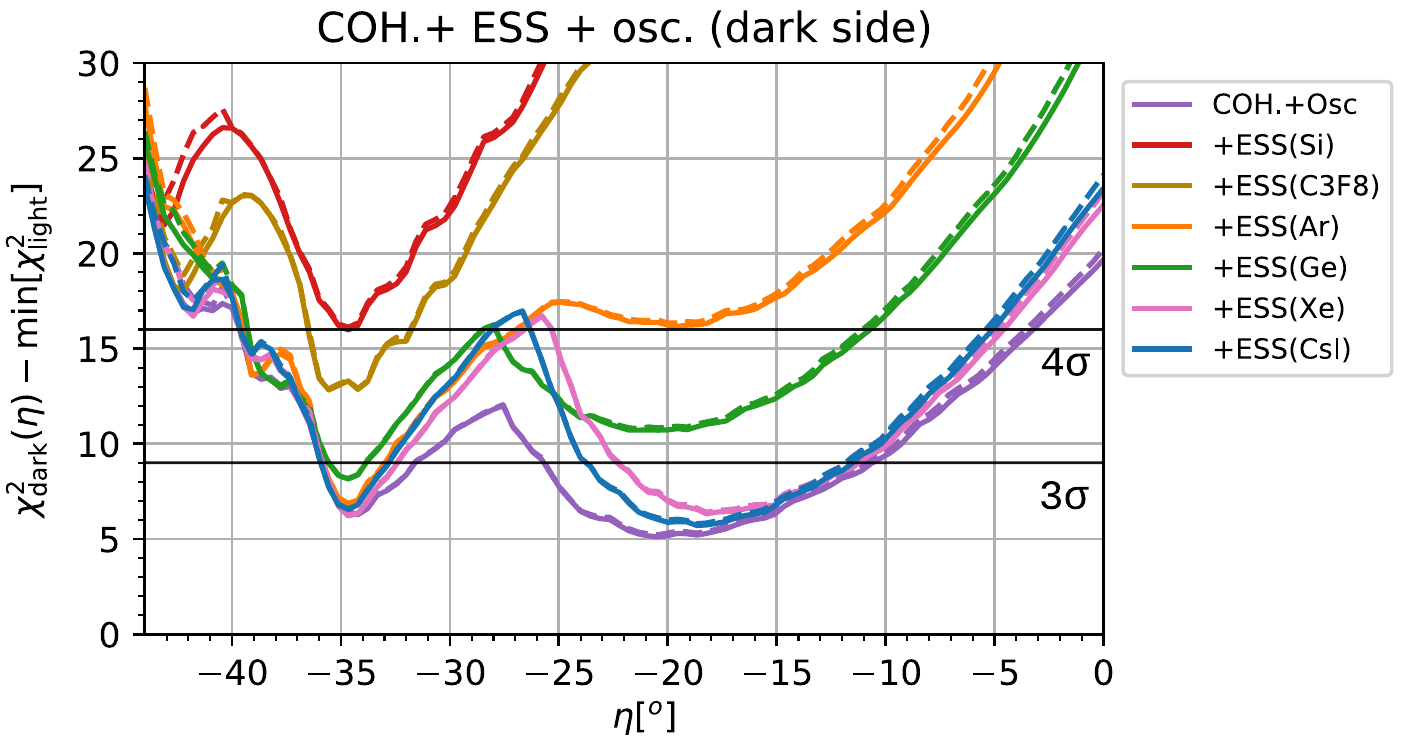}
    \mycaption{Expected sensitivity to exclude LMA-dark by a CE$\nu$NS measurement at ESS using different target materials.
    For dashed curves the off-diagonal 
    $\epsilon_{\alpha\beta}^\eta$ are fixed at zero, for solid curves we minimize with respect to them.
   }
    \label{fig:chi-eta-ess}
\end{figure}

In fig.~\ref{fig:chi-eta-ess} we show the sensitivity to exclude the LMA-dark solution from adding an ESS measurement with different target nuclei to present data. We see that Ar, Xe, CsI can lead only to a rather marginal improvement, increasing the $\Delta\chi^2$ by about 1 unit, and similar also for Ge, for which the improvement is about 3 units.
These nuclei have their blind spot close to $\eta \approx -35^\circ$ (c.f., tab.~\ref{tab:elements}), and therefore it is not possible to improve significantly around that value of $\eta$ with respect to the present situation. In contrast, a measurement using C$_3$F$_8$ and especially Si, can lead to a significant improvement. From tab.~\ref{tab:elements} we see that they have a neutron-to-proton ratio as well as $\eta_{\rm blind}$ sufficiently different from CsI, such that they will be able to exclude LMA-dark with $\Delta\chi^2 \approx 16.1$ (Si) and $13.0$ (C$_3$F$_8$).  

The complementarity of a Si measurement is illustrated also in figs.~\ref{fig:contours}
and \ref{fig:contours-off}. We see that in the relevant range of $\eta$ the ellipse from the Si measurement only marginally touches the LMA-dark band at the $3\sigma$ level. From these plots it is clear that also for the ESS measurement we observe a similar effect of off-diagonal NSI parameters as for COHERENT: they are negligible once constraints from oscillations are included, c.f.~dashed versus solid curves in fig.~\ref{fig:chi-eta-ess}.

\begin{figure}[t]
    \centering
    \includegraphics[scale=1]{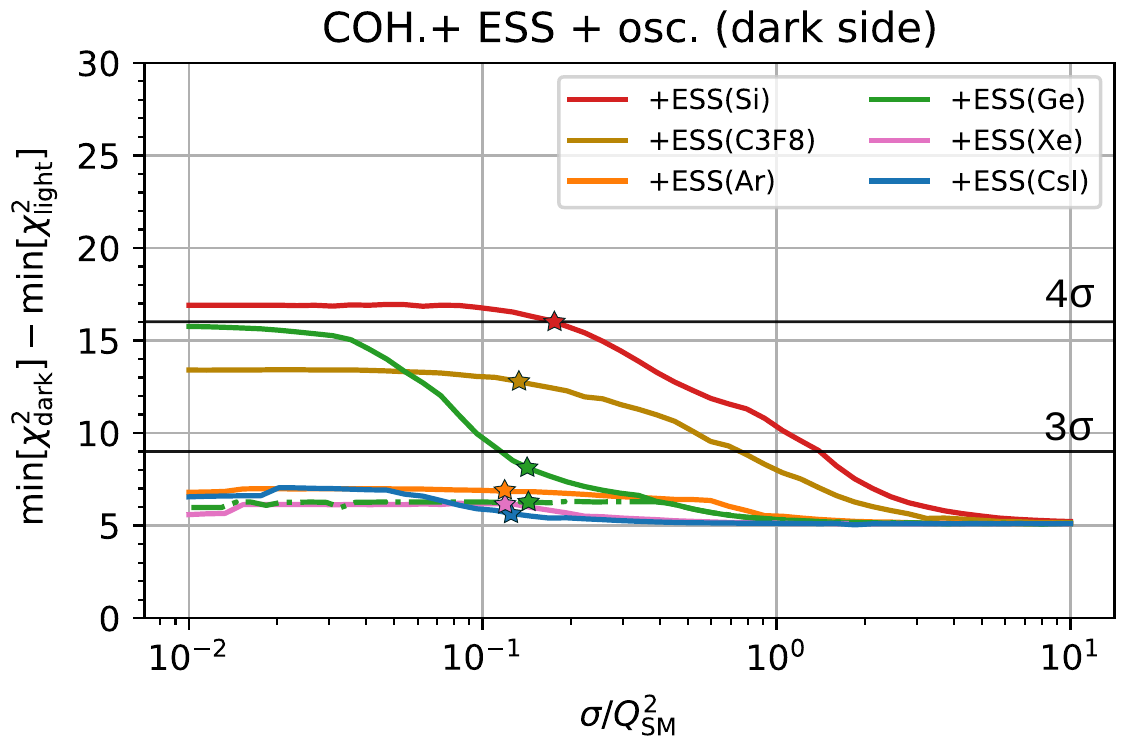}
    \mycaption{$\Delta\chi^2$ between the LMA-dark and LMA-light best-fit points as function of the relative measurement error $\sigma/Q^2_{\rm SM}$ for a CE$\nu$NS experiment at a stopped pion source for different target materials, assuming that the best-fit point corresponds to no NSI. 
    In scaling $\sigma$ we keep the ratio $\sigma_\mu/\sigma$ constant, using the value given in tab.~\ref{tab:elements}.
    The dashed-green curve shows the result for Ge for $\sigma_\mu \to\infty$. The stars indicate the assumptions for ESS sensitivities based on \cite{Baxter:2019mcx}. 
   }
    \label{fig:errorESS}
\end{figure}

In fig.~\ref{fig:errorESS} we address the question of which accuracy for a CE$\nu$NS measurement at a stopped pion source will be needed in order to disfavour LMA-dark significantly. We adopt the $\chi^2$ from eq.~\eqref{eq:chi-ess}, add it to the one from the present data and show the difference between the $\chi^2$ minima in the dark and light sides as a function of the relative measurement uncertainty of the weak charge. We see that for Si (C$_3$F$_8$), already for a rough measurement of  $\sigma/Q_{\rm SM}^2 \approx 1\, (0.5)$, LMA-dark will be disfavoured at $\Delta\chi^2 > 9$. 

At small values of $\sigma$ the curves become flat. The asymptotic value for a given target material is determined by the size and location of the ring in the plane of $\epsilon^\eta_{ee}$ and $\epsilon^\eta_{\mu\mu}$ relative to the LMA-dark band.\footnote{The small decrease at small $\sigma$ for Ar, Xe, CsI results from the fact that the best-fit point in the light side from current data is not exactly at $\epsilon^\eta = 0$. By adding hypothetical ESS data assuming no NSI, also the light-side best-fit point changes slightly, leading to the small decrease in $\Delta\chi^2$ between dark and light sides.} We see that Ar, Xe, and CsI targets would not reach $3\sigma$ even for an ideal measurement. However, the asymptotic values for Si, Ge, C$_3$F$_8$, are roughly 17, 16, 12.5, respectively. If evaluated for 1~dof this would correspond to about $4\sigma$ for Si and Ge and $3.5\sigma$ for C$_3$F$_8$. For Si and C$_3$F$_8$ the asymptotic sensitivity is already achieved for $\sigma/Q^2_{\rm SM}$ around 10\%, and already our default assumptions for ESS are rather close to them, as indicated by the stars in fig.~\ref{fig:errorESS}.

\begin{figure}
    \centering
    \includegraphics{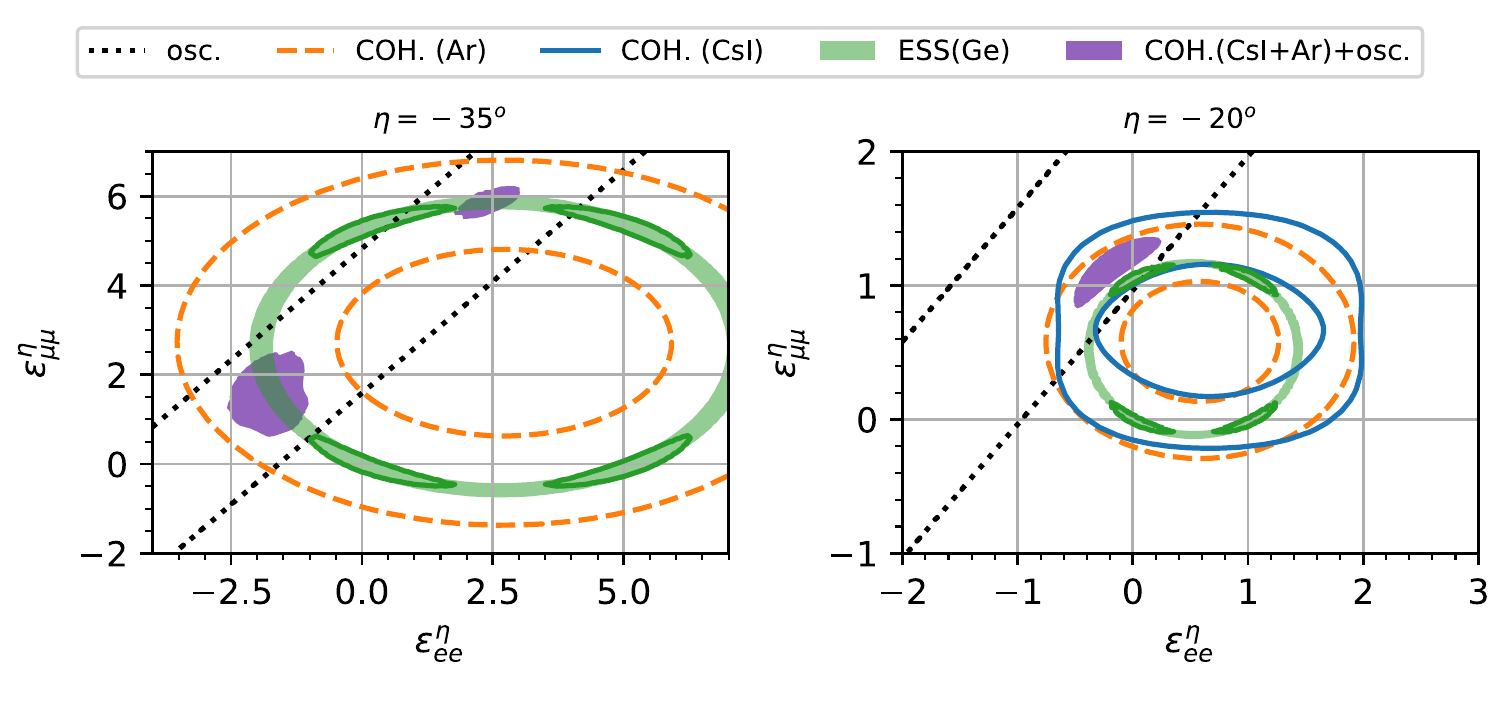}
    \mycaption{Same as fig.~\ref{fig:contours} but showing allowed regions at $3\sigma$ from a Ge target at a stopped pion source assuming a precision of $\sigma/Q^2_{\rm SM} = 0.028$, a factor 5 better than the ESS assumption from \cite{Baxter:2019mcx}. For the green-shaded region (green-solid contours) we assume $\sigma_\mu/\sigma = \infty \, (4.2)$.}
    \label{fig:ess-Ge-contours}
\end{figure}

Note that for Si and C$_3$F$_8$ we assume a total rate measurement, constraining only the combination $(Q_e^2/3 + 2 Q_\mu^2/3)$. For Ge we show in fig.~\ref{fig:errorESS} the impact of a partial separation of $Q_e^2$ and $Q_\mu^2$. The solid green curve corresponds to the situation where in addition to the total rate also $Q_\mu^2$ can be determined with a relative precision of $\sigma_\mu/\sigma = 4.2$, as motivated by the results of \cite{Baxter:2019mcx}. In contrast, the dashed-green curve shows the result for Ge using only the total rate, i.e., setting $\sigma_\mu\to\infty$. We see that for Ge the separate $Q_e^2$/$Q_\mu^2$ information is essential to disfavour LMA-dark at high significance. The reason for this becomes apparent in fig.~\ref{fig:ess-Ge-contours}, where we show a Ge measurement with a precision of a factor 5 better than the ESS assumption using total rate information only (green shaded). We see that the ring passes precisely through the two islands for $\eta\approx -35^\circ$ and therefore the degeneracy cannot be lifted for this value of $\eta$. The Ge constraint has a similar shape as the one from Ar, due to the similar value of the neutron-to-proton ratios, c.f.~tab.~\ref{tab:elements}. However, if in addition to the total rate also separate information on $Q^2_\mu$ is available, the ring becomes split into 4 islands (green-solid contours) and the degeneracy is resolved.

\subsection{CE$\nu$NS from reactor neutrinos}
\label{sec:reactors}

There are a number of ongoing or planned CE$\nu$NS experiments at nuclear reactors, see \cite{raimund_strauss_2020_4124156} for a
review. In this subsection we address the question of whether a reactor
measurement can also serve to resolve the LMA-dark degeneracy. 
The most relevant difference to pion sources is the pure electron flavour of the neutrino source. To be specific, we will consider as examples the ongoing CONNIE~\cite{Aguilar-Arevalo:2019jlr} and CONUS~\cite{Bonet:2020awv} experiments, which are using Si and Ge targets, respectively. Both experiments have published first results, which however, could not yet establish a significant measurement of CE$\nu$NS events. 

Similar as above, we estimate the sensitivity of future reactor experiments by assuming a determination of the weak charge. We define
\begin{equation}
    \chi^2_{\text{reac}}=\frac{(Q_{\rm SM}^2-Q^2_e)^2}
    {\sigma_{\rm reac}^2} \,, 
    \label{eq:chi-reactors}
\end{equation}
where again we assume that the best-fit point is at $Q_{\rm SM}^2$ and we adopt a measurement uncertainty of $\sigma_{\rm reac}/Q^2_{\rm SM} = 5\%$. While this appears to be a rather optimistic assumption, it serves to discuss the potential of a close-to-ultimate reactor measurement with respect to the LMA-dark ambiguity. 

\begin{figure}[t]
    \centering
    \includegraphics{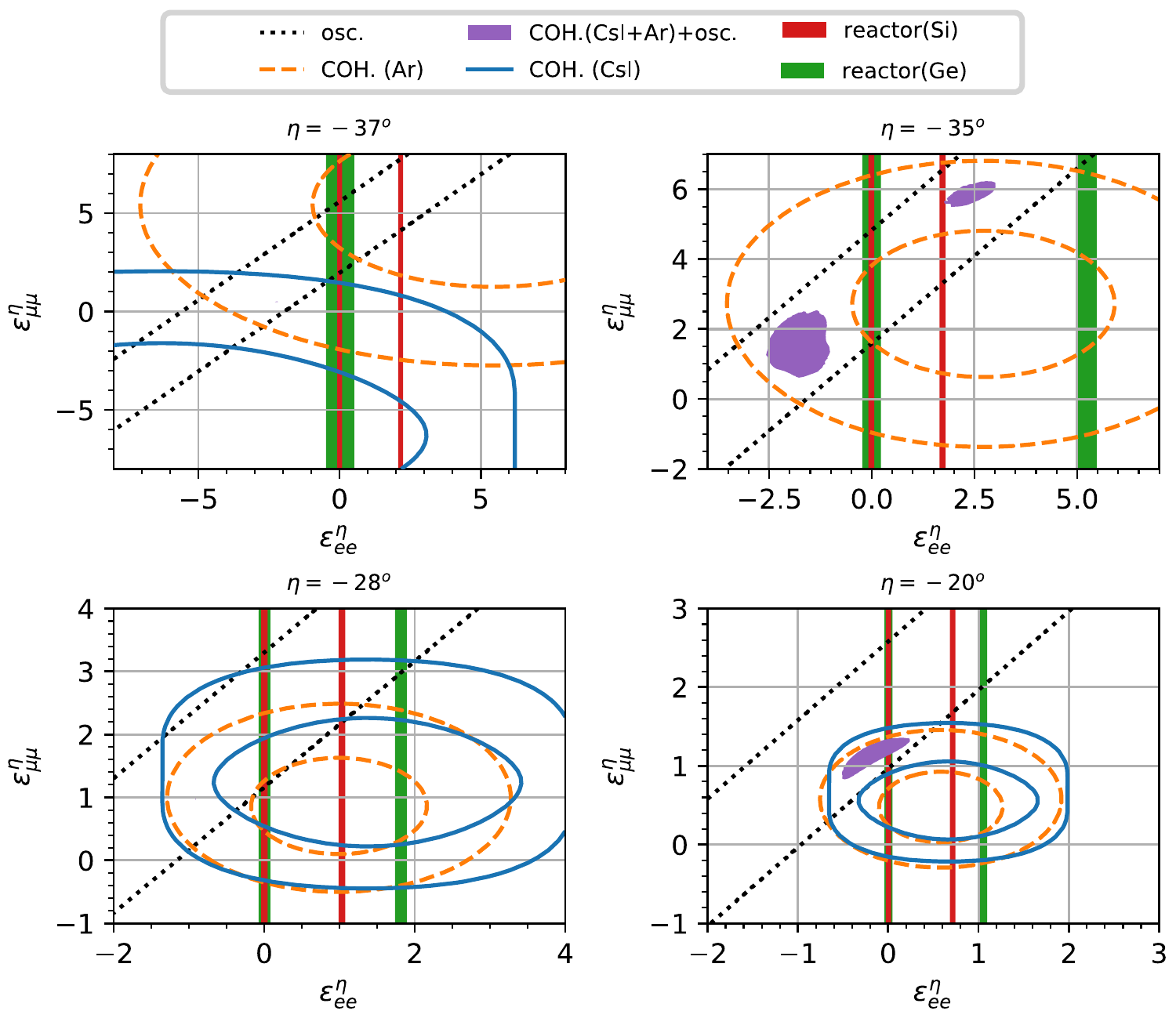}
    \mycaption{Same as fig.~\ref{fig:contours} with sensitivity of the reactor experiments using Ge and Si overlaid assuming a $5\%$ measurement of the weak charge $Q_{\rm SM}$.}
    \label{fig:contours+reactors}
\end{figure}

In fig.~\ref{fig:contours+reactors} we show the constraints from a reactor experiment in the plane of $\epsilon^\eta_{ee}$ and $\epsilon^\eta_{\mu\mu}$ for four values of $\eta$. Since they are sensitive only to $Q_e$ the limits are vertical bands in these plots. It is clear that for values of $\eta$, for which the LMA-dark allowed region overlaps with $\epsilon^\eta_{ee}=0$ such a measurement will not be able to exclude it. This is indeed the case for $\eta\approx -20^\circ$, as shown in the bottom-right panel of fig.~\ref{fig:contours+reactors}.

\begin{figure}[t]
    \centering
    \includegraphics[scale=0.87]{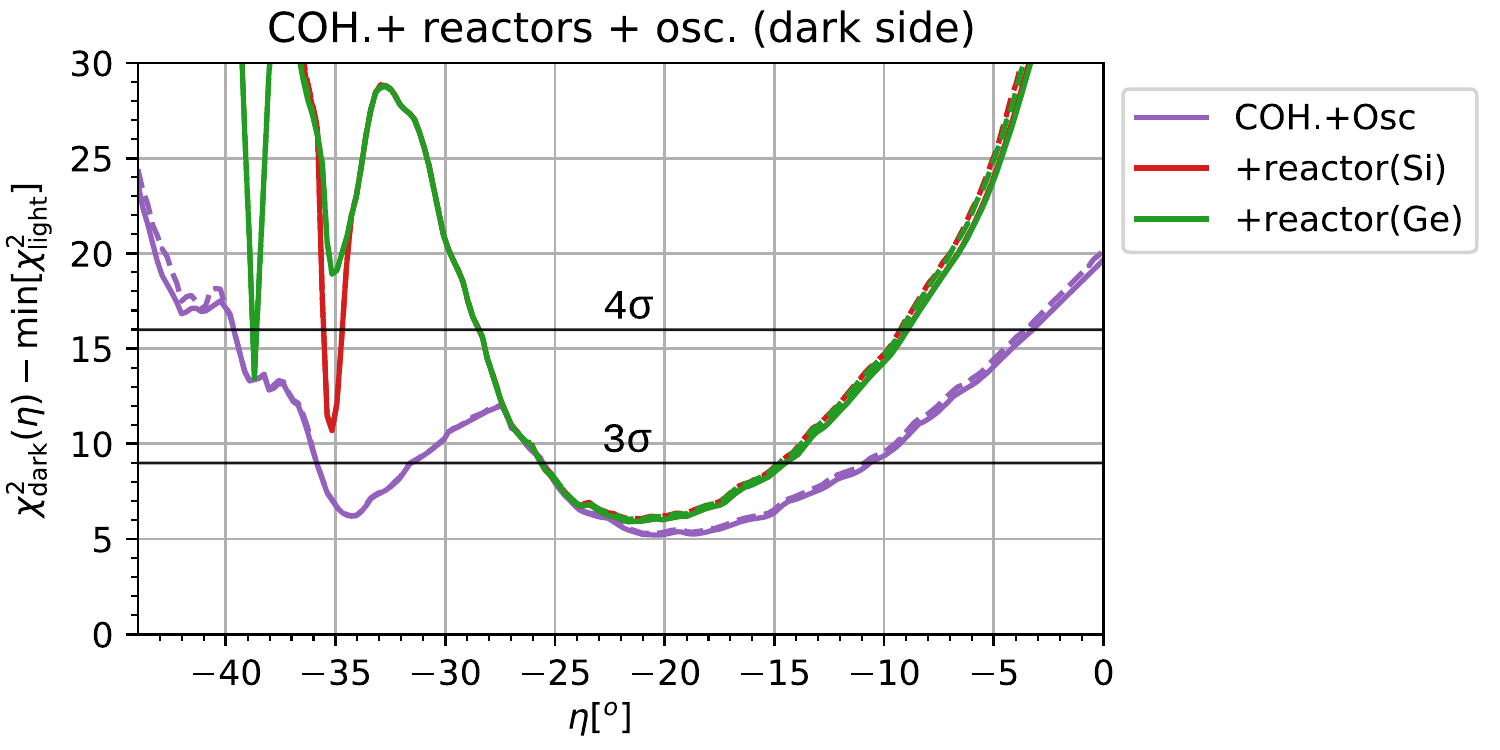}
    \mycaption{Sensitivity to exclude the LMA-dark solution by a hypothetical CE$\nu$NS measurement at a nuclear reactor using a Si (red curves) or a Ge (green curves) target. In both cases we assume a 5\% measurement of $Q_{\rm SM}$. For dashed curves off-diagonal $\epsilon^\eta_{\alpha\beta}$ are fixed to zero, for solid curves we minimize with respect to them. For $\eta \gtrsim -33^\circ$ the Si and Ge curves overlap.}
    \label{fig:chi-coh-osc-reactors}
\end{figure}

This behaviour is confirmed in fig.~\ref{fig:chi-coh-osc-reactors}, where the sensitivity to exclude the LMA-dark solution is shown as a function of $\eta$. 
We observe that for $-27^\circ < \eta \lesssim -15^\circ$ reactor experiments can essentially not improve with respect to the present situation. Some improvement is possible for $\eta < -27^\circ$. However, at certain values of $\eta$ in this region reactor experiments lose their sensitivity. The origin of this effect for Si is visible in the upper-right panel of  fig.~\ref{fig:contours+reactors}: it happens that the allowed band for 
non-zero $\epsilon^\eta_{ee}$ passes close to an island of the regions allowed by oscillations + COHERENT. The spike of the Ge experiment in fig.~\ref{fig:chi-coh-osc-reactors} has a similar origin. 

In fig.~\ref{fig:contours+reactors} the off-diagonal NSI parameters are fixed at zero. If we would allow them to vary freely, the region between the two vertical reactor bands would be filled, for a similar reason as discussed in sec.~\ref{sec:present} in the context of COHERENT. However, once the constraints from oscillation data are applied, the result is practically identical to the fixed case, c.f.~fig.~\ref{fig:chi-coh-osc-reactors}.

\begin{figure}[t]
    \centering
    \includegraphics[scale=0.9]{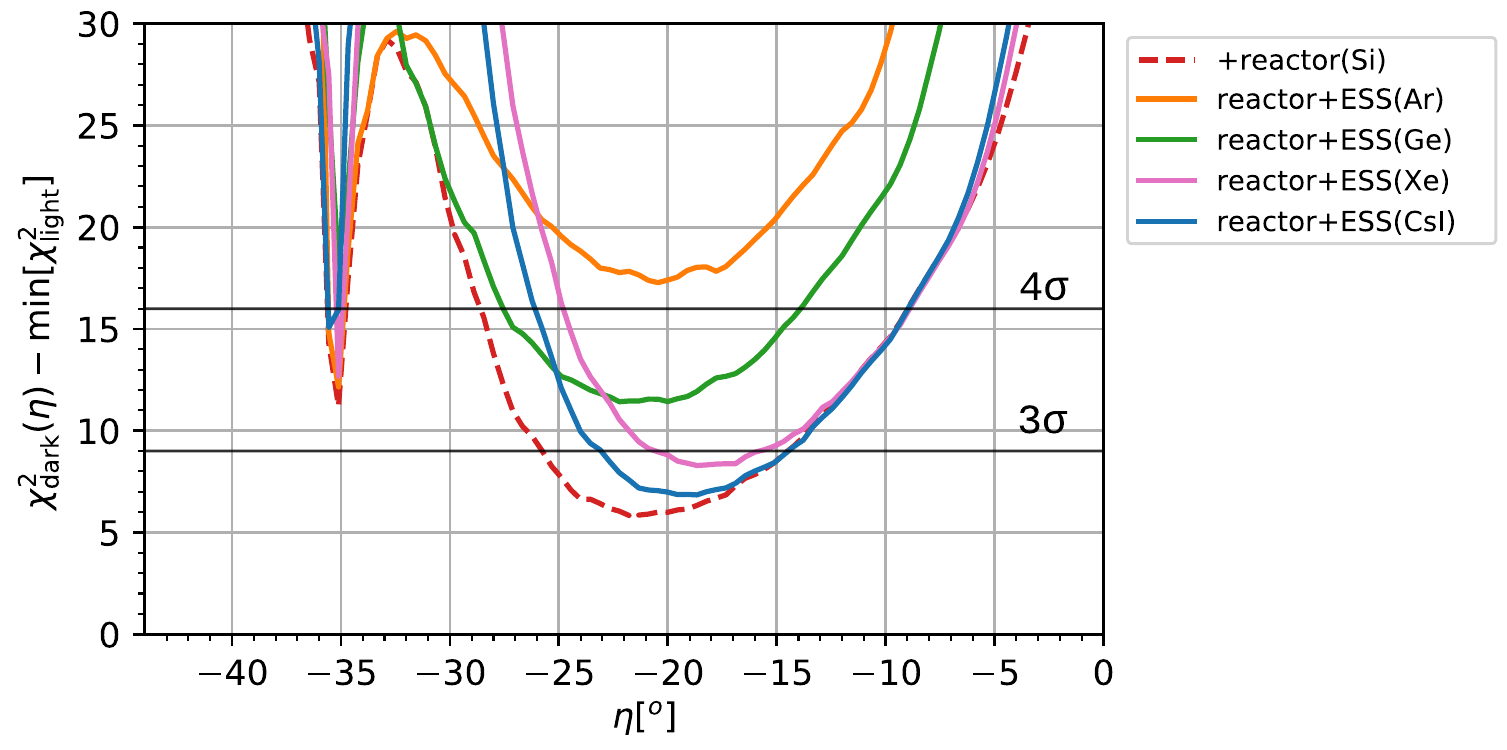}
    \mycaption{Sensitivity to exclude the LMA-dark degeneracy by a hypothetical CE$\nu$NS measurement at a nuclear reactor combined with ESS. The red-dashed curve corresponds to a Si target at a reactor. For the solid curves we combine reactor(Si) with measurements at the ESS assuming Ar, Ge, Xe, and CsI targets, see sec.~\ref{sec:ESS} for details.}
    \label{fig:chi-coh-osc-ESS-reactors}
\end{figure}

In fig.~\ref{fig:chi-coh-osc-ESS-reactors} we show some examples, where the complementarity of reactor and stopped pion source can be used to reach high significances. We combine a reactor measurement using Si with several target materials at ESS. (Results for using Ge at the reactor are very similar.) The ESS targets have been chosen such that by themselves they cannot reach $3\sigma$, c.f.~sec.~\ref{sec:ESS}. We observe that the combination of reactor with Ar (Ge) at ESS allows to reject LMA-dark at more than $4\sigma$ ($3\sigma$). For the heavy targets Xe and CsI a small region remains below $3\sigma$ around $\eta\approx -20^\circ$.

\section{Summary}
\label{sec:conclusion}

In this work we have updated the status of the LMA-dark degeneracy due to new data from the COHERENT experiment. We consider an NSI framework allowing for arbitrary relative couplings to up and down quarks parametrised by an angle $\eta$. Our simplifying assumptions are (i) that $\eta$ is flavour independent and (ii) we assume that flavour off-diagonal NSI coefficients are real. We find that the combination of data from oscillation experiments~\cite{Esteban:2018ppq} with latest COHERENT results~\cite{Akimov:2020pdx,coherent2020_CsI_update} disfavours LMA-dark with respect to LMA-light with $\Delta\chi^2 = 5$, i.e., at $2.2\sigma$, thanks to some complementarity in the latest CE$\nu$NS data on CsI and Ar. The degeneracy remains below $3\sigma$ in three islands in the parameter space, for $\eta$ in the intervals $[-35.9^\circ,-31.3^\circ]$ and $[-25.6^\circ,-10.5^\circ]$ and for flavour-diagonal NSI parameters of order one, see eq.~\eqref{eq:eps-LMA-D}.

We have investigated the potential of future CE$\nu$NS measurements to resolve the LMA-dark degeneracy at high confidence level. As an example we consider possible experiments at the ESS~\cite{Baxter:2019mcx} and we compare different detector materials. We find that light targets, with neutron-to-proton ratios $\approx 1$ are needed to exclude LMA-dark for all values of $\eta$. For example, a measurement on a Si or C$_3$F$_8$ target at the ESS, with a relative precision in the 10\% to 20\% range would exclude LMA-dark with a $\Delta\chi^2 = 16.1$ and 13.0, respectively, for arbitrary values of $\eta$. 
For these target materials, already a rough measurement of $\sigma/Q_{\rm SM}^2 \approx 50\%$ will disfavour LMA-dark at $\Delta\chi^2 > 9$. 
In contrast, a Ge target would require a precision below 10\% as well as a partial separation of the $\nu_e$ and $\nu_\mu$ flux contributions to the CE$\nu$NS rate to achieve a highly significant rejection of the degeneracy. 

A CE$\nu$NS measurement at a nuclear reactor will not be able to reject the LMA-dark degeneracy for any value of $\eta$, since only electron neutrinos are available at such a neutrino source. The reason is that the solution around $\eta\approx -20^\circ$ is consistent with $\epsilon^\eta_{ee} = 0$ and therefore predicts a CE$\nu$NS signal corresponding to the Standard Model at experiments using only the electron flavour. However, reactor experiments can contribute in certain target combinations of measurements at stopped pion sources to resolve the degeneracy.

\bigskip

The LMA-dark degeneracy implies a neutrino mass and mixing pattern qualitatively different from the standard scenario. It makes the determination of the neutrino mass ordering by oscillation experiments impossible and implies new physics contributing to neutrino interactions of similar size as weak interactions. Resolving this degeneracy is an essential prerequisite for neutrino physics to enter the precision era.

\subsection*{Acknowledgement}

We thank the authors of ref.~\cite{Esteban:2018ppq} for providing $\chi^2$-tables from their updated global oscillation analysis.
This project has received support from the European Union’s Horizon 2020 research and innovation programme under the Marie Sklodowska-Curie grant agreement No 860881-HIDDeN. This study was financed in part by the Coordenação de Aperfeiçoamento de Pessoal de Nível Superior -- Brasil (CAPES) -- Finance Code 001.  
MEC is grateful for the 140564/2018-7 funding from CNPQ.



\bibliographystyle{JHEP_improved}
\bibliography{./refs}

\end{document}